\documentclass{llncs}

\usepackage{url}
\usepackage{microtype}
\usepackage{booktabs} 
\usepackage{tikz}
\usetikzlibrary{arrows,shapes,positioning}
\usepackage[linesnumbered,ruled,vlined]{algorithm2e}
\usepackage{xspace}
\usepackage{subcaption}
\usepackage{xcolor}

\newcommand{\True}{\mathit{True}}

\newcommand{\fUnex}{f_{\mathit{Unexplored}}}
\newcommand{\remus}{ReMUS\xspace}

\begin{document}
\title{Recursive Online Enumeration of All Minimal Unsatisfiable Subsets}

\author{Jaroslav Bend\'ik \and Ivana \v Cern\'a \and Nikola Bene\v s}

\institute{Faculty of Informatics, Masaryk University, Brno, Czech Republic\\
\email{\{xbendik,cerna,xbenes3\}@fi.muni.cz}}

\authorrunning{J. Bend\'ik et~al.} 

\maketitle

\begin{abstract} 
In various areas of computer science,
we deal with a set of constraints to be satisfied. If the constraints cannot be satisfied simultaneously, it is desirable to identify the core problems among them. Such cores are called minimal unsatisfiable subsets (MUSes).
The more MUSes are identified, the more information about the conflicts among the constraints is obtained. However, a full enumeration of all MUSes is in general intractable due to the large number (even  exponential) of possible conflicts.
Moreover, to identify MUSes, algorithms have to test sets of constraints for their simultaneous satisfiability. The type of the test depends on the application domains. The complexity of the tests can be extremely high especially for domains like temporal logics, model checking, or SMT.

In this paper, we propose a recursive algorithm that identifies MUSes in an \emph{online} manner (i.e., one by one) and can be terminated at any time. The key feature of our algorithm is that it minimises the number of satisfiability tests and thus speeds up the computation. The algorithm is applicable to an arbitrary constraint domain and its effectiveness demonstrates itself especially in domains with expensive satisfiability checks.
We benchmark our algorithm against the state-of-the-art algorithm Marco on the Boolean and SMT constraint domains  and demonstrate that our algorithm really requires less satisfiability tests and consequently finds more MUSes in the given time limits.
 \end{abstract}


\section{Introduction}
In many different applications we are given a set of constraints (requirements) with the~goal to decide whether the set of constraints is satisfiable, i.e., whether all the constraints can hold simultaneously. In case the given set is shown to be unsatisfiable, we might be interested in an analysis of the unsatisfiability. Identification of minimal unsatisfiable subsets (MUSes) is a kind of such analysis. A minimal unsatisfiable subset is a set of constraints that are not simultaneously satisfiable, yet the elimination of any of them makes the set satisfiable.   We illustrate the problem on two different applications.


\newcommand{\rset}[3]{\{ \varphi_#1, \varphi_#2, \varphi_#3 \}}

In the  \textit{requirements analysis},   the constraints represent requirements on a~system that is being developed. Checking for satisfiability (also called \emph{consistency}) means checking whether all the requirements can be implemented at once. If the set of requirements is unsatisfiable, the~extraction of MUSes helps to identify and fix the conflicts among the requirements~\cite{bauch,DBLP:conf/issta/Bendik17}.

In some model checking techniques, such as the \emph{counterexample-guided abstraction refinement} (CEGAR)~\cite{cegar}, we are dealing with the following question: is the counterexample that was found in an abstract model feasible also in the concrete model? To answer this question, a formula $\mathit{cex} \wedge \mathit{conc}$ encoding both the counterexample $\mathit{cex}$ and the concrete model $\mathit{conc}$ is built and tested for satisfiability. If the formula is unsatisfiable, then the counterexample is \emph{spurious} and the negation of the formula $\mathit{cex} \wedge \mathit{conc}$ is used to refine the abstract model.
Since both $\mathit{cex}$ and $\mathit{conc}$ are often formed as a conjunction of smaller subformulas, the whole formula can be seen as a set of conjuncts (constraints). Andraus et al.~\cite{andraus2007cegar,cegar} found out that instead of using the negation of $\mathit{cex} \wedge \mathit{conc}$ for the refinement, it is better to identify MUSes of $\mathit{cex} \wedge \mathit{conc}$ and use
the negations of the MUSes to refine the abstract model.

Yet another applications of MUSes arise for example during formal equivalence checking~\cite{cohen2010designers}, proof based abstraction refinement~\cite{DBLP:conf/tacas/McMillanA03}, Boolean function bi-decomposition~\cite{bidecomposition}, circuit error diagnosis~\cite{han}, type debugging in Haskell~\cite{DBLP:conf/haskell/StuckeySW03}, or proof explanation in symbolic model checking~\cite{DBLP:conf/fmcad/GhassabaniWG17}.

The individual applications differ in the constraint domain. Perhaps the most widely used are Boolean and SMT constraints; these types of constraints arise for example in the CEGAR workflow. In the requirements analysis, the most common constraints are those expressed in a~temporal logic such as \emph{Linear Temporal Logic}~\cite{DBLP:conf/focs/Pnueli77} or \emph{Computational Tree Logic}~\cite{DBLP:conf/lop/ClarkeE81}.
The list of constraint domains in which MUS enumeration finds an application is quite long and new applications still arise. Therefore, we focus on MUS enumeration algorithms applicable in arbitrary constraint domains.

\smallskip
\noindent
{\bf Main contribution\ }
All algorithms solving the MUS enumeration problem have to get over two barriers. First,
the number of all MUSes can be exponential w.r.t. the number of constraints. Therefore, the complete enumeration of all MUSes can be intractable.  To overcome this limitation we present an algorithm for
\emph{online} MUS enumeration, i.e., an algorithm that enumerates MUSes one by one and can
be terminated at any time.

Second, algorithms for MUS enumeration need to test whether a given set of constraints is satisfiable. This  is typically a very hard task and it is thus desirable  to minimise the overall number of satisfiability queries.
To reduce the number of performed satisfiability queries, the majority of the state-of-the-art algorithms (for their detailed description see Section~\ref{sec:related-work}) try to exploit specific properties of particular constraint domains.  Most of the algorithms were evaluated only in the SAT domain (the domain of Boolean constraints). The SAT domain enjoys properties that can be used to significantly reduce the number of satisfiability queries and the state-of-the-art algorithms are thus very efficient in this domain. However, this might not be the case in domains for which no such domain specific properties exist.
Here, we present a novel algorithm that exploits both the domain specific as well as domains agnostics properties of the MUS enumeration problem.
First, the algorithm employs, as a black-box subroutine, a domain specific single MUS extraction algorithm which allows it to exploit specific properties of particular domains. Second, it recursively searches for MUSes in smaller and smaller subsets of the given set of constraints which allows it to directly reduce the number of performed satisfiability queries.

\section{Preliminaries and Problem Statement}
We are given a finite set of constraints $C$ with the property that each subset of $C$ is either \emph{satisfiable} or \emph{unsatisfiable}. The notion of satisfiability may vary in different constraint domains. The only assumption is that if a set $X$, $X \subseteq C$, is unsatisfiable, then all supersets of $X$ are   unsatisfiable as well. The sets of interests are defined as follows.

\begin{definition}[MUS, MSS, MCS]
Let $C$ be a~finite set of constraints and let $N \subseteq C$.
$N$ is a~\emph{minimal unsatisfiable subset} (MUS) of $C$ if $N$ is
unsatisfiable and $\forall c \in N : N \setminus \{c\}$ is satisfiable.
$N$ is a~\emph{maximal satisfiable subset} (MSS) of $C$ if $N$ is satisfiable
and $\forall c \in C \setminus N : N \cup \{c\}$ is unsatisfiable.
$N$ is a~\emph{minimal correction set} (MCS) of $C$ if $C \setminus N$ is a~MSS
of $C$.
\end{definition}
The maximality concept used  in the definition of a MSS is the set maximality and not the maximum cardinality as in the MaxSAT problem. Thus a constraint set $C$ can have multiple
MSSes with different cardinalities.

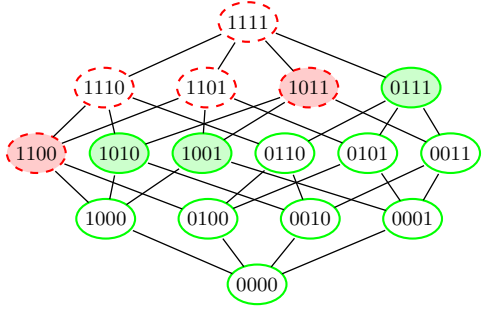
\begin{figure}[t!]
\centering
\begin{tikzpicture}[-,>=stealth', shorten >=1pt,auto,node distance=1.6cm,minimum size=0.62cm,
  thick,every node/.style={draw,white,text=black,ellipse,inner sep=0, outer sep=0, scale=0.85},
  sat/.style={draw,green,text=black},
  unsat/.style={dashed,red,text=black},
  mus/.style={fill=red!20},
  mss/.style={fill=green!20}]

\node[sat] (0) {$0000$};

\node[sat] (2) [above left = 0.5cm and 0.10cm of 0] {$0100$};
\node[sat] (1) [left of=2] {$1000$};
\node[sat] (3) [right of=2] {$0010$};
\node[sat] (4) [right of=3] {$0001$};

\node[unsat,mus] (12) [above left = 0.5cm and 0.37cm of 1] {$1100$};
\node[sat,mss] (13) [right = 0.32cm of 12] {$1010$};
\node[sat,mss] (14) [right = 0.32cm of 13] {$1001$};
\node[sat] (23) [right = 0.32cm of 14] {$0110$};
\node[sat] (24) [right = 0.32cm of 23] {$0101$};
\node[sat] (34) [right = 0.32cm of 24] {$0011$};

\node[unsat] (123) [above right = 0.5cm and 0.35cm of 12] {$1110$};
\node[unsat] (124) [right of = 123] {$1101$};
\node[unsat,mus] (134) [right of = 124] {$1011$};
\node[sat,mss] (234) [right of = 134] {$0111$};

\node[unsat] (1234) [above right = 0.5cm and 0cm of 124] {$1111$};

\path[every node/.style={font=\sffamily\small},line width=0.5pt]
(0) 	edge node [left,thin, dashed] {} (1)
		edge node [left] {} (2)
		edge node [left] {} (3)
		edge node [left] {} (4)
(1) 	edge node [left] {} (12)
		edge node [left] {} (13)
		edge node [left] {} (14)
(2) 	edge node [left] {} (12)
		edge node [left] {} (23)
		edge node [left] {} (24)
(3) 	edge node [left] {} (13)
		edge node [left] {} (23)
		edge node [left] {} (34)
(4) 	edge node [left] {} (14)
		edge node [left] {} (24)
		edge node [left] {} (34)

(12) 	edge node [left] {} (123)
		edge node [left] {} (124)
(13) 	edge node [left] {} (123)
		edge node [left] {} (134)
(14) 	edge node [left] {} (124)
		edge node [left] {} (134)
(23) 	edge node [left] {} (123)
		edge node [left] {} (234)
(24) 	edge node [left] {} (124)
		edge node [left] {} (234)
(34) 	edge node [left] {} (134)
		edge node [left] {} (234)

(123) 	edge node [left] {} (1234)
(124) 	edge node [left] {} (1234)
(134) 	edge node [left] {} (1234)
(234) 	edge node [left] {} (1234)
;
\end{tikzpicture}
\caption{Illustration of the power set of the set $C$ of constraints from the Example~1. We encode individual subsets of $C$ as bitvectors, e.g.~the subset $\{c_1,c_3,c_4\}$ is written as 1011. The subsets with dashed border are the unsatisfiable subsets and the others are satisfiable subsets. The MUSes and MSSes are filled with a~background colour.}
\label{ex:small_illustration}
\end{figure}

\begin{example}\label{ex:small}
 Assume that we are given a~set $C= \{c_1, c_2, c_3 , c_4 \}$ of four Boolean satisfiability constraints $c_1 = a$, $c_2 = \neg a$, $c_3 = b$, and $c_4 = \neg a \vee \neg b$.  Clearly, the whole set is unsatisfiable
as the first two constraints are negations of each other. There are two MUSes of $C$, namely
$\{c_1,c_2\}$, $\{c_1,c_3,c_4\}$. There are three MSSes of $C$, namely  $\{c_1,c_4\}$, $\{c_1,c_3\}$, and
$\{c_2,c_3,c_4\}$. Finally, there are three MCSes of~$C$, namely  $\{c_2,c_3\}$, $\{c_2,c_4\}$, and $\{c_1\}$. This example is illustrated in Fig.~\ref{ex:small_illustration}.
\end{example}

Another concept used in our work are the so-called \emph{critical constraints} which are defined as follows:
\begin{definition}[critical constraint]
Let $C$ be a~finite set of constraints and let $N \subseteq C$ be its unsatisfiable subset. A constraint $c \in N$ is a \emph{critical constraint} for $N$ if $N\setminus \{c\}$ is satisfiable.
\end{definition}
Note that $N$ is a MUS of $C$ if and only if each $c \in N$ is critical for $N$. Furthermore, if $c$ is a critical constraint for $C$ then $c$ has to be contained in every unsatisfiable subset of $C$, especially in every MUS of $C$.
Also, note that if $S$ is a MSS of $C$ and $\overline{S} = C \setminus S$ its complement (i.e. a MCS of $C$), then each $c \in \overline{S}$ is critical for $S \cup \{c \}$.

\begin{example}\label{ex:critical}
We illustrate the concept of critical constraints on two sets, $N$ and~$C$, where $C$ is the same set as in Example~\ref{ex:small}, and $N = C \setminus \{c_2\}$.
The constraint $c_1$ is the only critical constraint for $C$ whereas $N$ has three critical constraints: $c_1, c_3$, and~$c_4$.
\end{example}
\noindent
{\bf Problem Formulation\ }Given a set of constraints $C$,  enumerate all minimal unsatisfiable subsets of $C$ in an online manner while minimising  the number of constraints satisfiability queries. Moreover, we require an approach that is applicable to an~arbitrary constraint domain.

\section{Algorithm}
We start with some observations about the MUS enumeration problem and describe the main concepts   used in our algorithm.

The algorithm is given an unsatisfiable set of constraints $C$. To find all MUSes, the algorithm iteratively determines satisfiability    of   subsets of $C$. Initially, only the satisfiability of $C$ is determined and  at the end, satisfiability of all subsets of $C$ is determined. The algorithm maintains
a set \textit{Unexplored}  containing all subsets of $C$   whose satisfiability is not determined yet.
Recall that  if a set of constraints is satisfiable then all its subsets are   satisfiable as well. Therefore, if the  algorithm determines some $N \subseteq C$ to be satisfiable, then not only $N$ but also all of its subsets, denoted by $sub(N)$, become explored (i.e.~are removed from the set  \textit{Unexplored}
). Dually, if $N$ is unsatisfiable then  all of its supersets, denoted by $sup(N)$, are unsatisfiable and   become explored.

Since there are exponentially many subsets of $C$, it is intractable to represent the set \textit{Unexplored} explicitly.
Instead, we use a symbolic representation that is common in contemporary MUS enumeration algorithms~\cite{marco2,tome,DBLP:conf/fmcad/GhassabaniWG17}.
We encode $C = \{c_1, c_2, \ldots , c_n \}$ by using a set of Boolean variables $X = \{x_1, x_2, \ldots , x_n \}$. Each valuation of $X$ then corresponds to a subset of $C$. This allows us to represent the set of unexplored subsets $\mathit{Unexplored}$ using a Boolean formula $\fUnex$ such that each model of $\fUnex$ corresponds to an element of $\mathit{Unexplored}$. The formula is maintained as follows:
\begin{itemize}
	\item[$\bullet$] Initially $\fUnex = \mathit{True}$ since all of $\mathcal{P}(C)$ are unexplored.

	\smallskip
	\item[$\bullet$] To remove a satisfiable set $N \subseteq C$ and all its subsets from the set  \textit{Unexplored} we add to $\fUnex$ the clause $\bigvee_{i: c_i \not\in N} x_i$.

	\smallskip
	\item[$\bullet$] Symmetrically, to remove  an unsatisfiable set $N \subseteq C$ and all its supersets  from the set  \textit{Unexplored}  we add to $\fUnex$  the clause $\bigvee_{i: c_i \in N} \neg x_i$.
\end{itemize}
To get an element of $\mathit{Unexplored}$, we ask a SAT solver for a model of $\fUnex$.

\begin{example}\label{ex:unex}
Let us illustrate the symbolic representation on $C = \{ c_1, c_2, c_3 \}$. If  all subsets of $C$ are unexplored then $\fUnex = \True$. If   $\{c_1, c_3 \}$ is determined to be unsatisfiable and $\{c_1, c_2 \}$ to be satisfiable, then $\fUnex$ is updated to $\True \wedge (\neg x_1 \vee \neg x_3) \wedge (x_3)$.
\end{example}

One of the approaches  (see e.g.~\cite{marco,tome,emus}) how to find a MUS of $C$ is to find an unexplored unsatisfiable subset, called a \emph{seed}, and then use any algorithm that finds a MUS of the seed (this algorithm is often denoted as a \emph{shrink} procedure). In our algorithm we  use a black-box subroutine for shrinking. This allows us to always employ the best available, domain specific, single MUS extraction algorithm.

The key question is how to find an unexplored unsatisfiable subset (a seed).
Due to the monotonicity of the satisfiability of individual subsets (w.r.t. subset inclusion), satisfiable subsets are typically smaller and, dually, unsatisfiable subsets are more concentrated among the larger subsets. Therefore, we search for seeds among \emph{maximal unexplored subsets}. A~set $S_{max}$ is a maximal unexplored subset of $C$ iff $S_{max} \subseteq C$, $S_{max} \in \mathit{Unexplored}$, and each of the proper supersets of  $S_{max}$ is explored. The maximal unexplored subsets correspond to the \emph{maximal models} of $\fUnex$. Thus, in order to get a maximal unexplored subset $S_{max}$, we ask a SAT solver for such a model. If $S_{max}$ is unsatisfiable, we use it as a seed for the shrinking procedure and compute a MUS of $C$.

The idea of searching for seeds among the maximal unexplored subsets is already used in some contemporary MUS enumeration algorithms~\cite{marco2,emus,tome,DBLP:conf/fmcad/GhassabaniWG17}. However, none of the algorithms specify which maximal unexplored subset should be used for finding a seed. They just ask a SAT solver for an arbitrary maximal model of $\fUnex$ (maximal unexplored subset). We found that the choice of maximal unexplored subset to be used is very important. The complexity of the shrinking procedure, in general, depends on the cardinality (the number of constraints) of the seed. Thus, an ideal option would be to search for a seed among the maximal unexplored subsets with the minimum cardinality, i.e. to find a \emph{minimum maximal model} of $\fUnex$. However, finding such a model is very expensive, especially compared to finding an arbitrary maximal model of $\fUnex$. In order to find an arbitrary maximal model, we can just instruct the SAT solver to use $\mathit{True}$ as a default polarity of variables during solving (this can be done e.g.~in the miniSAT~\cite{minisat} solver). On the other hand, finding a minimum maximal model of $\fUnex$ is a hard optimisation problem.

\begin{figure}[!t]
\centering
\begin{subfigure}{.5\textwidth}
  \centering
  \includegraphics[scale=0.35]{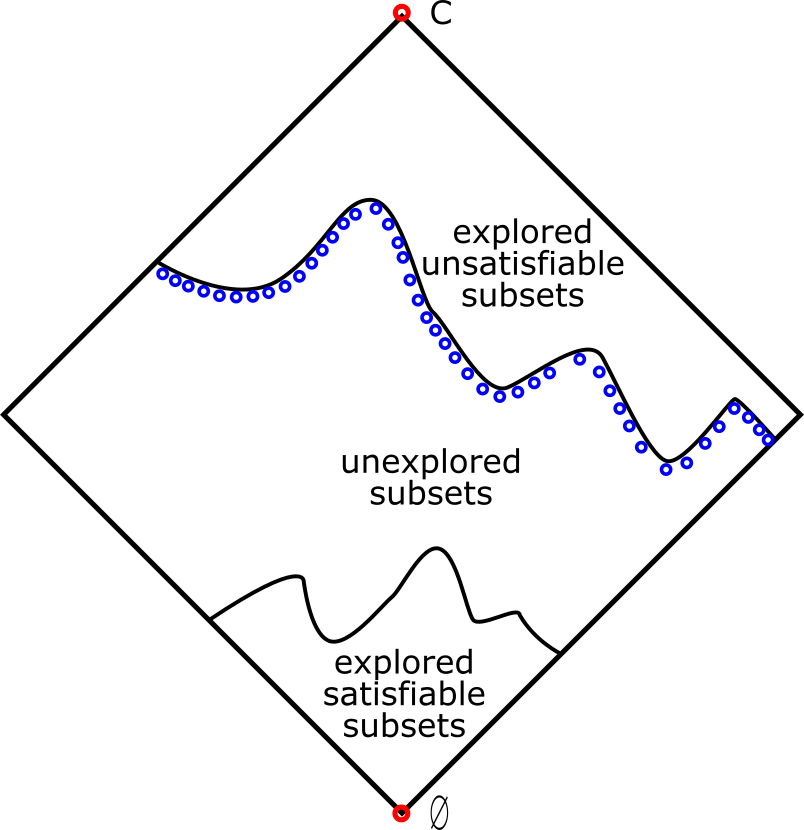}
  \caption{}
  \label{cube-distribution}
\end{subfigure}%
\begin{subfigure}{.5\textwidth}
  \centering
  \includegraphics[scale=0.35]{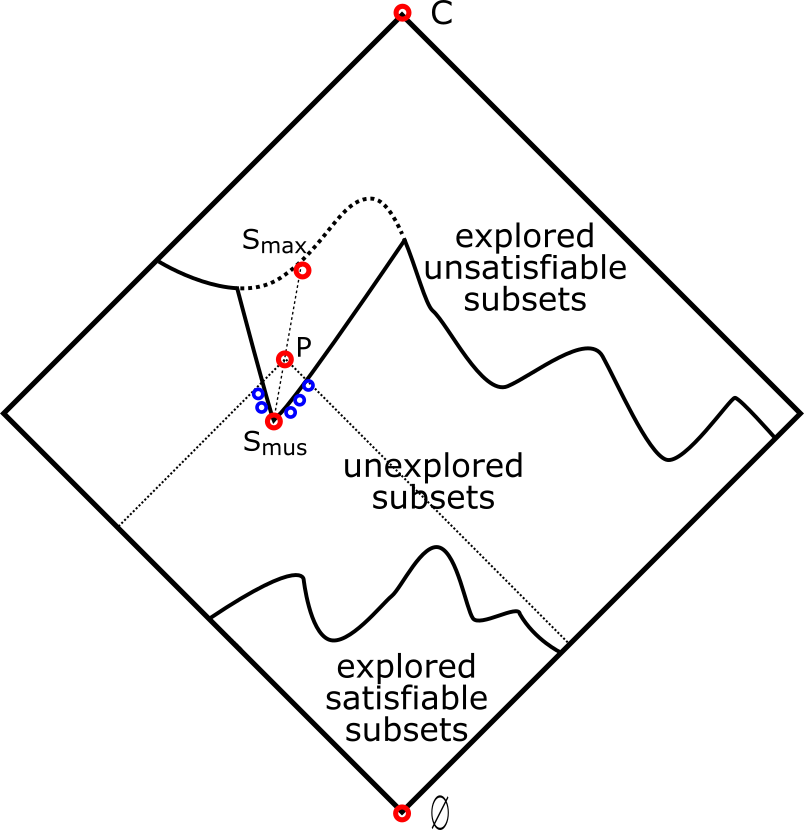}
  \caption{}
  \label{cube-recursion}
\end{subfigure}\caption{Illustration of our seed selection approach. Figure~\ref{cube-distribution} illustrates the division of subsets of $C$ into explored satisfiable, explored unsatisfiable, and unexplored subsets. The blue circle nodes represent the maximal unexplored subsets of $C$. Figure~\ref{cube-recursion} shows a previously used seed $S_{max}$, a MUS $S_{mus}$ that was found based on the seed, a set $P$ such that $S_{mus} \subseteq P \subseteq S_{max}$, and maximal unexplored subsets of $P$ (blue circle nodes).}
\label{cubes}
\end{figure}

We propose a way of finding seeds that are relatively small, yet cheap to be found.
To find the first seed we are repeatedly asking the SAT solver for an arbitrary maximal unexplored subset of $C$ until we obtain some unsatisfiable maximal unexplored subset $S_{max}$. Then, we use $S_{max}$ as the first seed and shrink it to the first MUS $S_{mus}$.
In order to find the next seed, we use a more sophisticated approach. Instead of searching for a seed among the maximal unexplored subsets of $C$, we restrict the search space so that the next seed is smaller than the previous one.
In particular, we choose a~set $P$ such that $S_{mus} \subseteq P \subseteq S_{max}$ and search for the new seed among the maximal unexplored subsets of $P$.
Note that the maximal unexplored subsets of $P$ do not have to be maximal unexplored subsets of $C$.
Furthermore, $P$ is necessarily unsatisfiable and all seeds found within it are necessarily smaller than the previous seed $S_{max}$ since $P \subseteq S_{max}$.
By choosing the next seed among the maximal unexplored subsets of $P$, we de facto solve the problem in a~recursive manner. Instead of finding a new MUS of $C$, we find a MUS of $P$, which is necessarily also a MUS of $C$. Each next seed is found based on the previous one, i.e. we keep to recursively reducing the search space as far as we can. Once we end up in a subset $P$ of $C$ such that the whole $\mathcal{P}(P)$ is explored, we backtrack from the recursion.
The approach is illustrated in Fig.~\ref{cubes}.

\smallskip
In our algorithm we also use   critical constraints. For mining critical constraints we use the maximal unexplored subsets that are satisfiable.  Every satisfiable maximal unexplored subset $S_{max}$ of $C$ is a maximal satisfiable subset (MSS) of $C$ as every  superset of $S_{max}$ is explored. If it were satisfiable then due to monotonicity $S_{max}$ should also be explored (which it is not).
Thus, for every $c \in C \setminus S_{max}$ it holds that $S_{max} \cup \{c\}$ is unsatisfiable and $c$ is a critical constraint for $S_{max} \cup \{c\}$.

The critical constraints are used in two different situations. The first situation arises when we are searching for a seed and the selected maximal unexplored subset $S_{max}$ of $C$ is   satisfiable. In such a case we can try to pick another maximal unexplored subset of $C$ and check it for satisfiability.
 However, for reasons mentioned above, we try to search for small seeds. Therefore we recursively search for a maximal unexplored  subset of $S_{max} \cup \{c\}$, where $c$ is a critical constraint for $S_{max} \cup \{c\}$. The set $S_{max} \cup \{c\}$ is definitely unsatisfiable and its  cardinality  is not greater than the cardinality of $C$.

Many modern shrinking algorithms \cite{DBLP:conf/cav/BacchusK15,muser,DBLP:conf/fmcad/NadelRS13}   use critical constraints to speed up their computation. Every MUS of $C$ has to contain all the critical constraints for~$C$ and this helps the shrinking procedure to narrow the search space. The critical constraints  for  $C$ have to be delivered to the shrinking algorithm together with the seed. Our algorithm can compute and accumulate critical constraints very effectively even when recursively traversing the space.  If  $X$ and $Y$ are two sets of constraints such that $X \supseteq Y $, then every critical constraint for $X$ is also a critical constraint for $Y$. The algorithm thus can utilise the known critical constraints even in its recursive part.

\subsection{Description of the Algorithm}
The pseudocode of our algorithm is shown in Algorithm~\ref{alg-basic}. The computation of the algorithm starts with an initialisation phase  followed by a call of the procedure \texttt{FindMUSes}, which is the core procedure of our algorithm.

The procedure \texttt{FindMUSes} has two input parameters: $S$ and $\mathit{criticals}$.  $S$ is an unsatisfiable set of constraints and the  procedure outputs     MUSes of $S$. The set $\mathit{criticals}$  contains (currently known) critical constraints for $S$ and is used for the shrinking procedure.   In each iteration, the procedure \texttt{FindMUSes} picks a maximal unexplored subset $S_{\mathit{max}}$ of $S$ and tests it for satisfiability.
If $S_{\mathit{max}}$ is satisfiable, then it is guaranteed to be a MSS of $S$. Thus, the complement $S_{mcs} = S \setminus S_{max}$  of $S_{max}$ is an MCS of $S$ and it can be used to obtain critical constraints. If $|S_{mcs}| = 1$, then the single constraint that forms $S_{mcs}$ is guaranteed to be a critical constraint for $S$ and it is thus added into $\mathit{criticals}$. Otherwise, the procedure recursively calls itself on $(S_{max} \cup \{c\}, \mathit{criticals} \cup \{c\})$  for each $c \in S_{mcs}$ since each such $c$ is guaranteed to be a critical constraint for $ S_{max} \cup \{c\}$.

In the other case, when $S_{\mathit{max}}$ is unsatisfiable, then   $S_{max}$  is shrunk
to a MUS $S_{\mathit{mus}}$ (note that the set $\mathit{criticals}$ of critical constraints is provided to the shrinking procedure). The newly computed  $S_{\mathit{mus}}$ is used to reduce the dimension of the space in which another MUSes are searched for. Namely,  the procedure picks   some $P$, $S_{mus} \subset P \subset S_{max}$, and recursively calls itself on $ P$. After the recursive call terminates, the procedure continues with the next iteration.

 The main idea behind the recursion is to search for MUSes of sets smaller than $S$ and thus lower the complexity of performed operations.  Naturally, there is a trade-off between the complexity of operations and the expected number of MUSes occurring in the chosen subspace and thus it might be very tricky to  find an optimal   $P$. In our algorithm we choose $P$ so that $|P| = 0.9\cdot |S_{\mathit{max}}|$, where $|P|$ and $|S_{\mathit{max}}|$ are cardinalities of the two sets, respectively. We form $P$  by adding a corresponding number of constraints from $S_{\mathit{max}}$ to $S_{\mathit{mus}}$. Note that it might happen that $|S_{\mathit{mus}}| \geq 0.9 \cdot |S_{\mathit{max}}|$; in such a case the algorithm skips the recursion call and continues with the next iteration.

\begin{algorithm}[t!]
\DontPrintSemicolon
\SetKwInOut{Input}{input}\SetKwInOut{Output}{output}
\SetKwFunction{Init}{Init}
\SetKwFunction{FindMUSes}{FindMUSes}
\SetKwProg{Fn}{Function}{:}{}
\SetKwFunction{Shrink}{Shrink}

\Fn{\Init{$C$}}{
\Input{an unsatisfiable set of constraints $C$}
	$\mathit{Unexplored} \gets \mathcal{P}(C)$\;
	\FindMUSes{$C$, $\emptyset$}\;
}

\setcounter{AlgoLine}{0}
\Fn{\FindMUSes{$S$, $\mathit{criticals}$}}{
	\While{$\mathit{Unexplored} \cap \mathcal{P}(S) \neq \emptyset$}{
		$S_{max} \gets$ a maximal unexplored subset of $S$\;
		\eIf{$S_{max}$ is satisfiable}{
			$\mathit{Unexplored} \gets \mathit{Unexplored} \setminus Sub(S_{max})$\;
			$S_{mcs} \gets S \setminus S_{max}$\;
			\eIf{$|S_{mcs}| = 1$}{
				$\mathit{criticals} \gets \mathit{criticals} \cup S_{mcs}$\;			
			}{
				\For{\textbf{\emph{each}} $c \in S_{mcs}$}{
					\FindMUSes{$S_{max} \cup \{c\}$, $\mathit{criticals} \cup \{c\}$}\;
				}
			}
		}{
			$S_{mus} \gets $\Shrink{$S_{max}$, $\mathit{criticals}$}\;
			\textbf{output} $S_{mus}$\;
			$\mathit{Unexplored} \gets \mathit{Unexplored} \setminus (Sup(S_{mus}) \cup Sub(S_{mus}))$\;
			\If{$|S_{mus}| < 0.9 \cdot |S_{max}|$}{
				$P \gets$ subset such that $S_{mus} \subset P \subset S_{max}$, $|P| = 0.9 \cdot |S_{max}|$\;
				\FindMUSes($P$, $\mathit{criticals}$)\;			
			}
		}
	}
}	 
\caption{\remus}
\label{alg-basic}
\end{algorithm}

The set \textit{Unexplored} is  updated appropriately during the whole computation. Note that the set \textit{Unexplored} is shared among the individual recursive calls; in particular if the algorithm determines some subset $S$ to be unsatisfiable then all of its supersets (w.r.t. the original search space) are deduced to be unsatisfiable. On the other hand, the maximal unexplored subsets (and their complements) are local and are defined with respect to the current search space.

\smallskip
\noindent
{\bf Correctness\ }
The algorithm outputs only the results of \emph{shrinking} which is assumed to be a correct MUS extraction procedure. Each MUS is produced only once since only unexplored subsets are shrunk and each MUS is removed from the set ${\mathit{Unexplored}}$ immediately after producing.
Only subsets whose status is known are removed from the set ${\mathit{Unexplored}}$ thus no MUS is excluded from the computation.
The algorithm terminates and all MUSes are found since the size of ${\mathit{Unexplored}}$ is reduced after every iteration.

\section{Related Work}\label{sec:related-work}
The list of existing approaches to the MUS enumeration problem is short, especially compared to the amount of work   dealing with   a single MUS extraction~\cite{dmuser,DBLP:conf/fmcad/BelovM11,muser,DBLP:conf/sat/BelovS11,nadel,haifa}.
Moreover, existing  algorithms for the MUS enumeration are tailored mainly to Boolean constraints~\cite{gasca,gleeson,DBLP:conf/cpaior/BacchusK16,DBLP:conf/cav/BacchusK15} and cannot be applied to other constraints. The approaches that focus on MUS enumeration in general constraint systems can be divided into two categories: approaches that  compute MUSes directly and those that rely on the \emph{hitting set duality}.

\medskip
\noindent
\textbf{Direct MUS enumeration}  \\
The early algorithms were based on explicit enumeration of every subset of the unsatisfiable constraint system.
As far as we know, the MUS enumeration was pioneered by Hou~\cite{hou} in the field of diagnosis. Hou's algorithm checks every subset for satisfiability starting with the whole set of constraints and exploring its power set in a tree-like structure. Also, some pruning rules that allow skipping irrelevant branches are presented. This approach was revisited and further improved by Han and Lee~\cite{han} and by de la Banda et al.~\cite{banda}. Another approach using step-by-step powerset exploration was recently proposed by Bauch et al.~\cite{bauch}. The authors of this work focus on constraints expressed using LTL formulas; however, their algorithm can be used for any type of constraints. Explicit exploration of the power set is the bottleneck of all of the  above mentioned algorithms as the size of the power set is exponential to the number of constraints in the system.

Liffiton et al.~\cite{marco} and Silva et al.~\cite{emus} developed independently two nearly identical algorithms: MARCO~\cite{marco} and eMUS~\cite{emus}.
Both algorithms were later merged and presented~\cite{marco2} under the name of MARCO.
Among the existing MUS enumeration algorithms, MARCO is perhaps the one most similar to ReMUS.
It uses symbolic representation of the power set and is able to produce MUSes incrementally during its computation in a relatively steady rate.
In order to find individual MUSes, it iteratively picks  maximal unexplored subsets of the original set of constraints and checks them for satisfiability. The unsatisfiable subsets are shrunk, using a black-box procedure, into MUSes. Contrary to ReMUS, MARCO does not tend to reduce the size of the sets to be shrunk and thus to directly reduce the number of performed satisfiability checks. Instead, it assumes that the black-box shrinking procedure would do the trick. MARCO is very efficient in constraint domains for which efficient shrink procedures exist. However, in the other domains, it is very inefficient. Especially due the the fact that it shrinks the maximal unexplored subsets of the original set of constraints, i.e. it shrinks (relatively) very large sets.

In our previous work~\cite{tome}, we have presented the algorithm TOME that also produces MUSes in an online manner. It iteratively uses binary search to find the so-called local MUSes/MSSes. Each local MUS/MSS is optionally, based on its size (cardinality), shrunk/grown to a global MUS/MSS. TOME tries to predict the complexity of performing the shrinking/growing procedure and only those shrinks/grows that seem to be easy to perform are actually performed.
TOME is very efficient in constraint domains for which no efficient shrinking and growing procedure exist.
On the other hand, in domains like Boolean constraints, the effort needed to find local MUSes and MSSes outweighs the effort needed to perform the shrinks and grows.

\medskip
\noindent
\textbf{Hitting set duality based approaches} \\
There is a well known relationship between MUSes and MCSes  based on the concept of  \emph{hitting sets}.
Given a collection $\Omega$ of sets, a hitting set $H$ for $\Omega$ is a~set such that $\forall S \in \Omega: H \cap S \not= \emptyset$. A hitting set is called \emph{minimal} if none of its proper subsets is a hitting set.  If $C$ is a set of constraints and $N \subseteq C$, then the \emph{minimal hitting set duality}~\cite{duality} claims that $N$ is a MUS of $C$ iff $N$ is a minimal hitting set of the set of all the MCSes of $C$.

The hitting set duality is used  for example in CAMUS~\cite{camus} and DAA~\cite{daa}. CAMUS works in two phases. It first computes all MCSes of a given constraint set and then   finds all MUSes by computing all minimal hitting sets of these MCSes. A significant shortcoming of CAMUS is that the first phase can be intractable as the number of MCSes can be exponential in the number of constraints and all MCSes must be found before the first MUS can be produced.

The algorithm DAA~\cite{daa} is able to produce some MUSes before the enumeration of MCSes is completed. DAA starts each iteration with computing a minimal hitting set $H$ of currently known MCSes and tests $H$ for satisfiability. If $H$ is unsatisfiable, it is guaranteed to be a~MUS. In the other case, $H$ is \emph{grown} into a~MSS whose complement is a~MCS, i.e.~the set of known MCSes is enlarged. As in the case of MARCO, DAA can use any existing algorithm for a~single MSS/MCS extraction to perform the grow.

MARCO, CAMUS and DAA were experimentally compared in the Boolean constraints domain~\cite{marco2} and CAMUS has shown to be the fastest in enumerating all MUSes in the tractable cases. However, in the intractable cases, MARCO was able to produce at least some MUSes, while CAMUS often got stuck in the phase of MCSes enumeration. DAA was much slower than CAMUS in the case of complete MUSes enumeration and also slower than MARCO in the case of partial MUS enumeration. The main drawbacks of DAA are the complexity of computing minimal hitting sets and no guarantee on the rate of~MUS~production.

\medskip
\noindent
Bacchus and Katsirelos proposed a MUS enumeration algorithm called MCS-MUS-BT~\cite{DBLP:conf/cpaior/BacchusK16} which is also based on recursion and uses MCSes to extract critical constraints. However, the algorithm is tailored for the SAT domain and, thus, cannot be applied in an arbitrary constraint domain. Moreover, the computation of MCSes is an integral part of MCS-MUS-BT, and the MCSes are computed in a different way taking up to linearly many satisfiability checks to compute each MCS. MCS-MUS-BT does not use black-box shrinking procedures and the recursion is not driven by previously found MUSes.

\section{Implementation}
We implemented ReMUS into a publicly available tool\footnote{https://www.fi.muni.cz/~xbendik/remus/}. The tool currently supports three different constraint domains: SAT (Boolean constraints), SMT, and LTL. It employs several external tools. In particular, it uses the SAT solver miniSAT~\cite{minisat} for maintaining $\fUnex$, and miniSAT is also used as a satisfiability solver in the SAT domain. 
The tools z3~\cite{z3} and SPOT~\cite{spot} are used as satisfiability solvers in the SMT and LTL domains, respectively. Moreover, our tool uses the single mus extractor muser2~\cite{muser} as a black-box shrink subroutine in the SAT domain. In the other domains, we use our custom implementation of the shrinking procedures. 

\section{Experimental Evaluation}
Here, we report results of our experimental evaluation. Besides evaluating ReMUS, we also provide a comparison with the latest tool implementation\footnote{\url{https://sun.iwu.edu/~mliffito/marco/}} of the state-of-the-art MUS enumeration algorithm MARCO~\cite{marco2}. 
The comparison is done in the SAT and SMT domains since these are the domains supported by the MARCO tool. 
Note that MARCO uses same external procedures as ReMUS, i.e. a satisfiability solver, a shrinking procedure, and a SAT solver for maintaining unexplored subsets. All these external procedures are implemented in the MARCO tool same as in the ReMUS tool, i.e. using miniSAT~\cite{minisat}, z3~\cite{z3}, and muser2~\cite{muser}.

There are three main criteria for the comparison: 
1) the number of output MUSes within a given time limit, 2) the number of satisfiability checks required to output individual MUSes, and 3) the time required to output individual MUSes.

\subsection{Benchmarks and Experimental Setup}
The experiments in the SAT domain were conducted on a collection of 292 Boolean CNF benchmarks that were taken from the MUS track of the SAT 2011 competition\footnote{\url{http://www.cril.univ-artois.fr/SAT11/}}.
The benchmarks range in their size from 70 to 16 million constraints and use from 26 to 4.4 million variables.
This collection of benchmarks has been already used in several papers that focus on the problem of MUS enumeration, see e.g.~\cite{tome,marco,marco2,camus}.
In the SMT domain, we used a set of 433 benchmarks that were used in the work by Griggio et al.~\cite{griggio}.
The benchmarks were selected from the library SMT-LIB\footnote{\url{http://www.smt-lib.org/}}, and include instances from the QF\_UF, QF\_IDL, QF\_RDL, QF\_LIA and QF\_LRA divisions.
The size of the benchmarks ranges from 5 to 145422 constraints.

The experiments were
run on an Intel(R) Xeon (R) CPU E5-2630 v2, 2.60GHz, 125 GB memory machine
running Arch Linux 4.9.40-l-lts. All experiments were run using a time limit of 3600 seconds. Complete results are available at \mbox{\url{https://www.fi.muni.cz/~xbendik/remus/}}.

\section{Experimental Results}

\begin{figure}[!t]
\centering
\begin{subfigure}{.5\textwidth}
  \centering
  \includegraphics[scale=0.5]{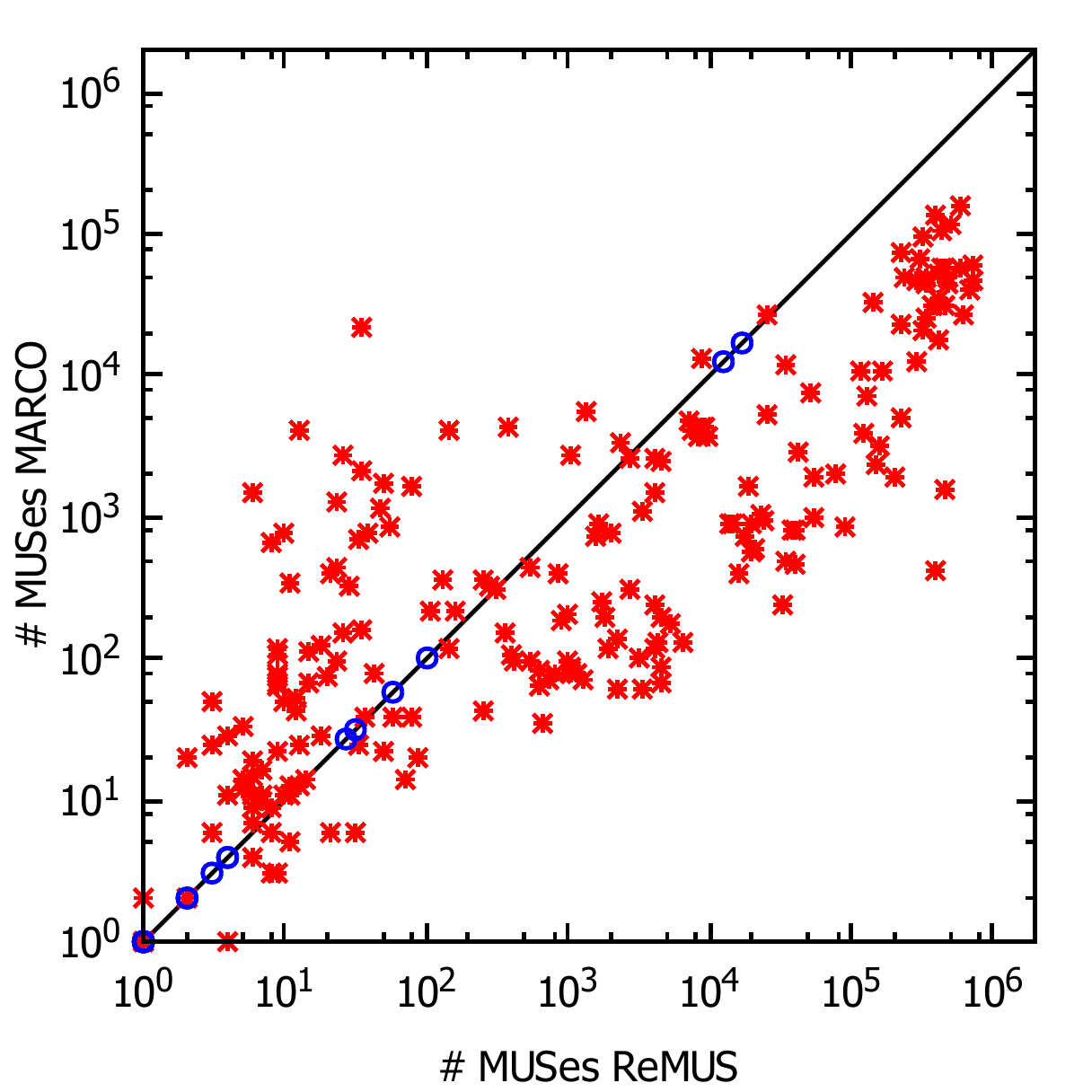}
  \caption{SAT domain.}
  \label{res:scatter_sat}
\end{subfigure}%
\begin{subfigure}{.5\textwidth}
  \centering
  \includegraphics[scale=0.5]{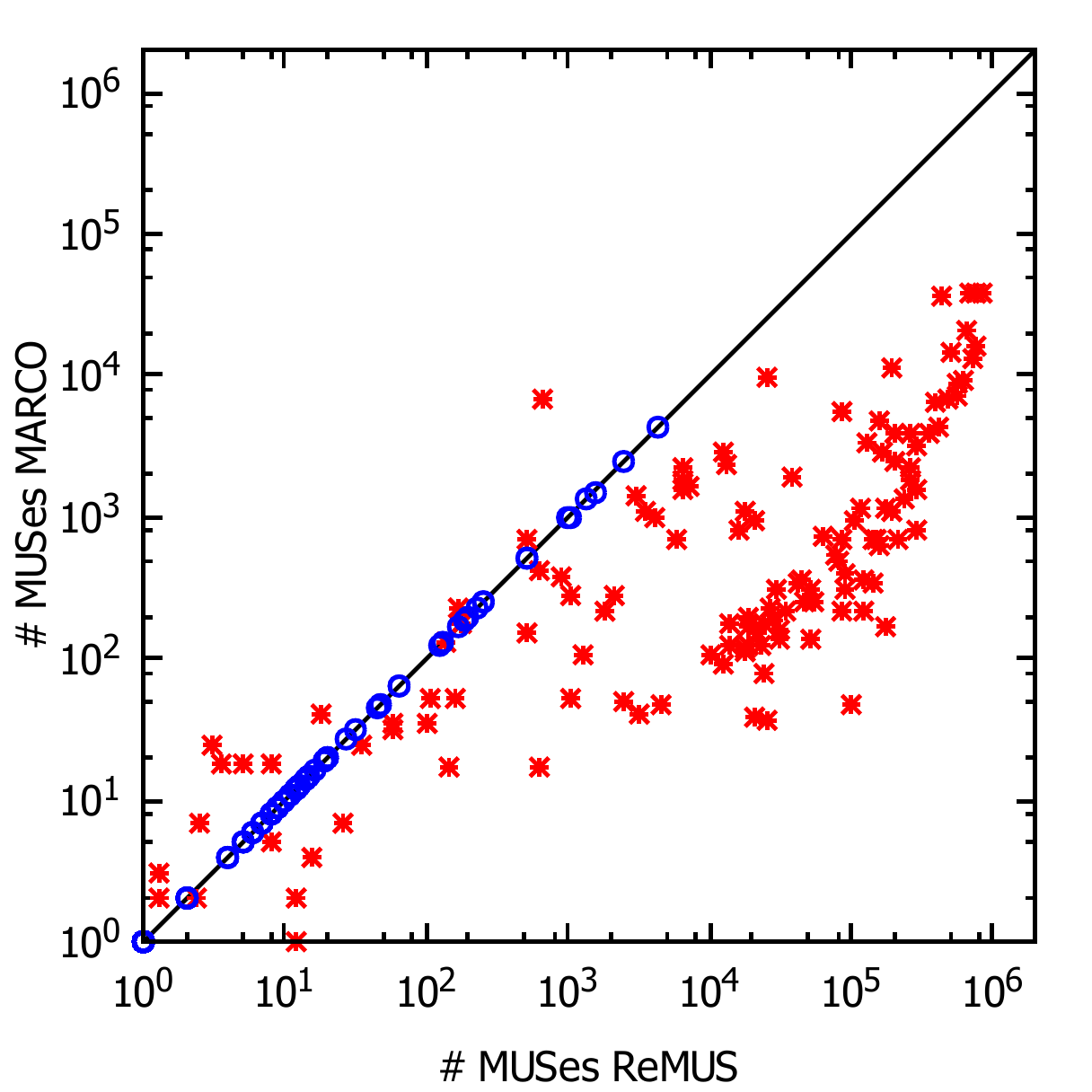}
  \caption{SMT domain.}
  \label{res:scatter_smt}
\end{subfigure}
\caption{Scatter plots comparing the number of produced MUSes. Blue points represent the benchmarks where both algorithms finished the computation.}
\label{res:scatters}
\end{figure}

\subsection{Number of Output MUSes}
In this section, we examine the performance of evaluated algorithms in terms of number of produced MUSes within the given time limit of 3600 seconds.
Due to the potentially exponentially many MUSes in each instance, the
complete MUS enumeration is generally intractable. Moreover, even producing
a~single MUS can be intractable for larger instances as it naturally includes solving
the satisfiability problem, which is relatively hard to solve in the SAT and SMT domains. 
Within the given time limit, only in 216 SAT and 238 SMT instances both algorithms found more than two MUSes. 
Only in 24 SAT and 245 SMT instances both algorithms finished the computation.

Figure~\ref{res:scatters} provides scatter plots that compare both evaluated algorithms on individual
benchmarks in the SAT and SMT domains.
Each point in the plot represents the result achieved by the two compared
algorithms on one particular instance; one algorithm determines the position on the vertical
axis and the other one the position on the horizontal axis. 
MARCO found strictly more MUSes than ReMUS in 76 SAT and 15 SMT instances. On the other hand, ReMUS found strictly more MUSes than MARCO in 162 SAT and 118 SMT instances. Note that in the SMT domain, ReMUS was often better than MARCO by two orders of magnitude.

\subsection{Performed Checks per MUS}
\begin{figure}[!t]
\centering
\begin{subfigure}{.5\textwidth}
  \centering
  \includegraphics[scale=0.5]{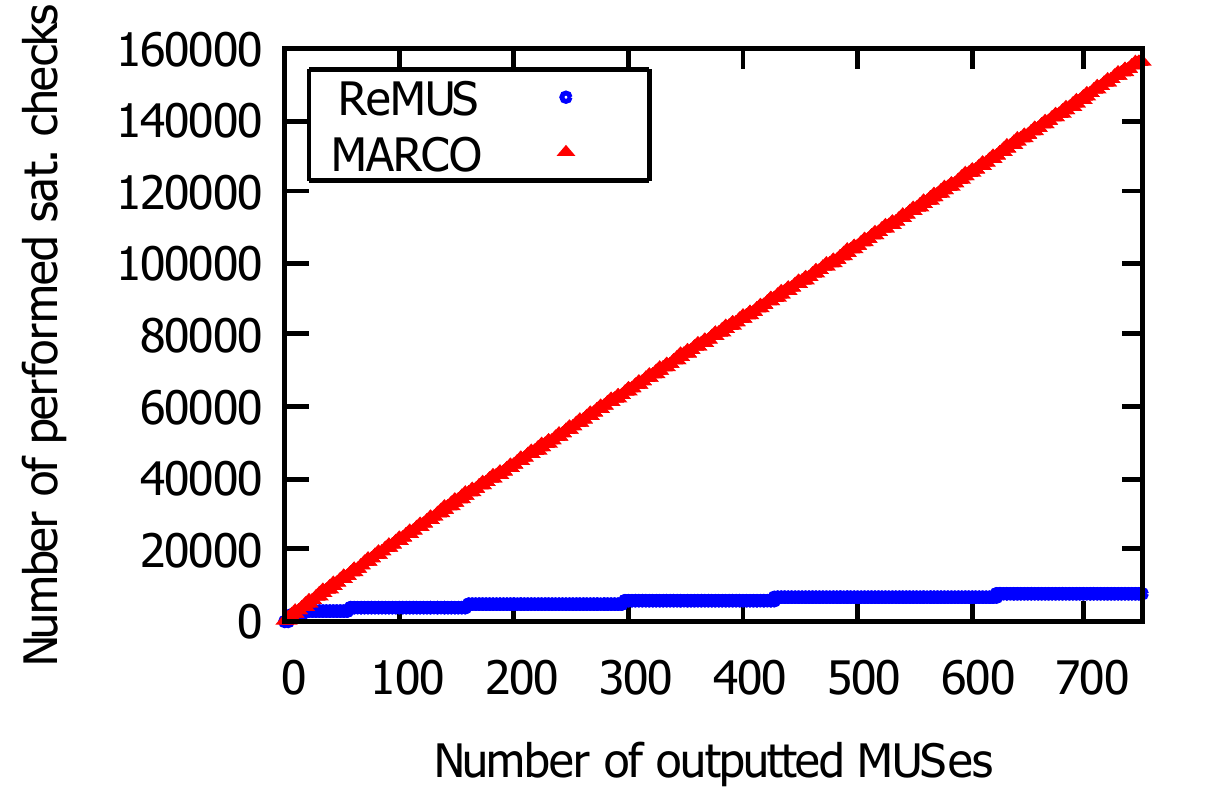}
  \caption{SAT domain.}
  \label{res:checks_sat}
\end{subfigure}%
\begin{subfigure}{.5\textwidth}
  \centering
  \includegraphics[scale=0.5]{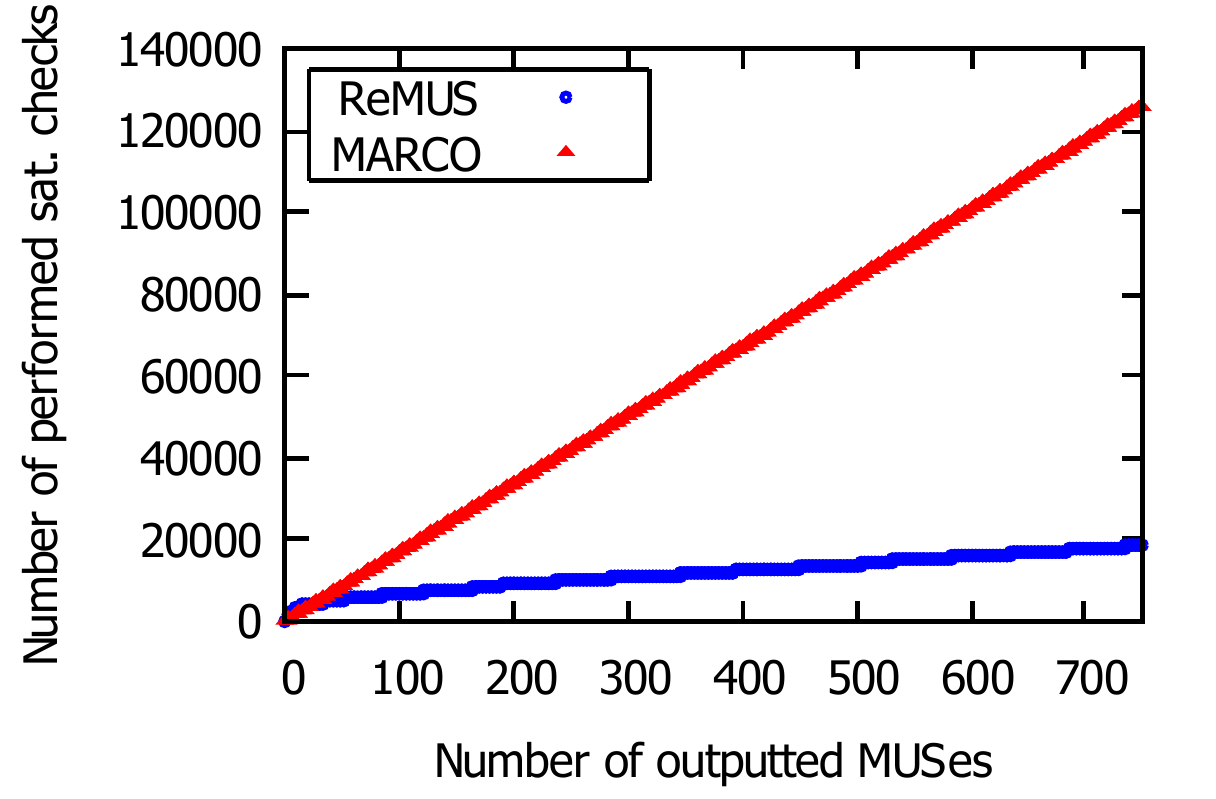}
  \caption{SMT domain.}
  \label{res:checks_smt}
\end{subfigure}
\caption{Plots showing the average number of performed satisfiability checks required to output individual MUSes.}
\label{res:checks}
\end{figure}

In this section, we focus on the main optimisation criterion of our algorithm: the number of checks required to output individual MUSes.
This number differs for different benchmarks since individual benchmarks vary in many aspects such as the size of the benchmarks and the size of the MUSes contained in the benchmarks. Therefore, we focus on average values.
Plots in Fig.~\ref{res:checks} show the average number of performed satisfiability checks required to output the first 750 MUSes. A point with coordinates $(x,y)$ states that the algorithm needed to perform $y$ satisfiability checks on average in order to output the first $x$ MUSes. We used only a subset of the benchmarks to compute the average values since only for some benchmarks both algorithms found at least 750 MUSes. In particular, 70 and 51 benchmarks were used to compute the average values in the SAT and SMT domains, respectively.

ReMUS is clearly superior to MARCO in the number of satisfiability checks required to output individual MUSes. 
This happens both due to the fact that ReMUS gradually, in a recursive way, reduces the dimension of the search space (and thus shrink smaller seeds), as well as due the fact that ReMUS mines and accumulates critical constraints to speed up the shrinking procedures.

\subsection{Elapsed Time per MUS}
The fact that ReMUS 
requires less satisfiability checks than MARCO to output
individual MUSes does not necessarily mean that it is also faster than MARCO in producing individual MUSes.
The time spent by ReMUS to maintain the recursive calls
while trying to save some satisfiability checks might not be worth it if the checks are easy to perform. We need to answer a domain specific question: is the 
price of performing 
satisfiability checks high enough?

\begin{figure}[!t]
\centering
\begin{subfigure}{.5\textwidth}
  \centering
  \includegraphics[scale=0.5]{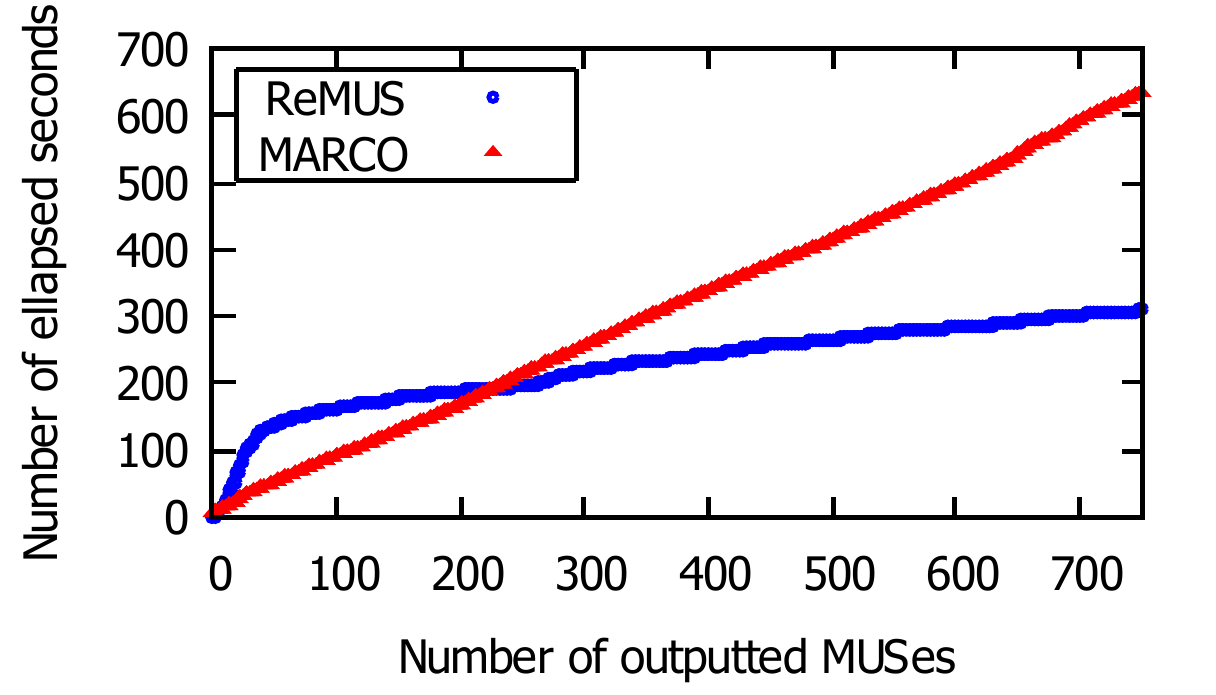}
  \caption{SAT domain.}
  \label{res:time_sat}
\end{subfigure}%
\begin{subfigure}{.5\textwidth}
  \centering
  \includegraphics[scale=0.5]{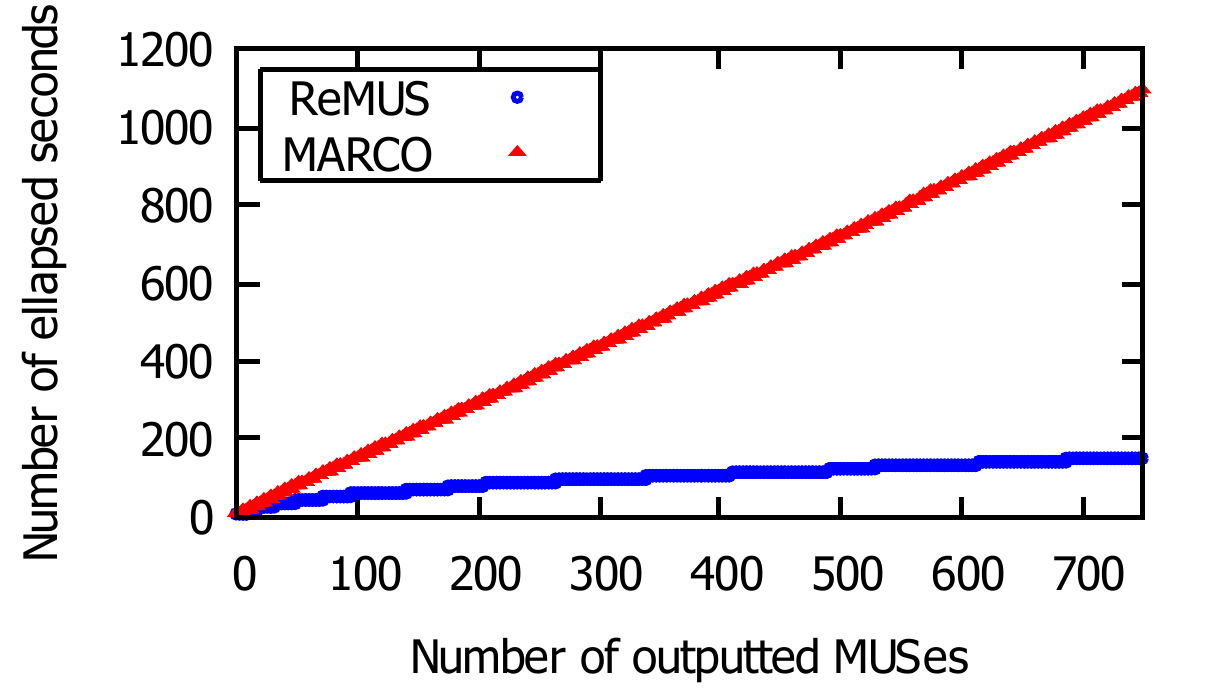}
  \caption{SMT domain.}
  \label{res:time_smt}
\end{subfigure}
\caption{Plots showing the average amount of time required to output individual~MUSes.}
\label{res:time}
\end{figure}

To answer this question for the SAT and SMT domains, we took the 70 SAT and 51 SMT benchmarks in which both algorithms produced at least 750 MUSes and computed the average amount of time required to output individual MUSes. 
The results are shown in Fig.~\ref{res:time}.
A point with coordinates $(x,y)$ states that in order to output the first $x$ MUSes the algorithm required $y$ seconds on average.
In the SMT domain, ReMUS is significantly faster from the very beginning of the computation. In the SAT domain, MARCO is faster during the first three minutes, yet afterwards ReMUS becomes much faster. 

\subsection{Evaluation}
Experimental results demonstrate that ReMUS outperformed MARCO on almost all the SMT instances 
and on a majority of the SAT instances. However, on some SAT instances, ReMUS was quite struggling, especially at the beginning of the computation.  Here, we point out three characteristics of benchmarks/domains that affect  the performance of ReMUS.

First, ReMUS tends to minimise the number of performed satisfiability checks. Therefore the higher is the complexity of the satisfiability checks, the more is the tendency to minimise the number of performed checks worth it. 
Second, the motivation behind finding small seeds for shrinking procedure is based on the fact that, in general, the larger the seed is, the more satisfiability checks are required. However, some constraint domains might enjoy domain specific properties that allow to shrink the seed very efficiently, regardless of the size of the seed. In particular, the CNF form of Boolean (SAT) formulas allows to significantly reduce the number of performed satisfiability checks~\cite{muser,haifa,dmuser}. 
Finally, the reduction of the search space is driven by previously found MUSes. In order to perform deep recursion calls, the input instance has to contain many MUSes. Moreover, there have to be some similar MUSes, i.e. there has to be a subset that is relatively small and yet contains several MUSes. 

In the SMT domain, the shrinking procedures are currently not so advanced as in the SAT domain, and the complexity of the satisfiability checks in the SMT domain is often larger than in the SAT domain. 
Thus, even a small reduction of the size of the seeds leads to a notable improvement in the overall efficiency.
On the other hand, in the SAT domain, either a significant reduction of the size of the seeds (i.e. deep recursion calls) or 
a large number of cumulated critical constraints
is required to speed up the shrinking.

\section{Conclusion}
We have presented the algorithm ReMUS for online enumeration of MUSes that is applicable to an arbitrary constraint domain.
We observed that the time required to output individual MUSes generally correlates with the number of satisfiability checks performed to output the MUSes. The novelty of our algorithm lies in exploiting both the domain specific as well as domain agnostic properties of the MUS enumeration problem to reduce the number of performed satisfiability checks, and thus also reduce the time required to output individual MUSes.
The main idea of the algorithm is to recursively search
for MUSes in smaller and smaller subsets of a given set of constraints.
Moreover, the algorithm cumulates critical constraints and uses them to speed up single MUS extraction subroutines.
We have experimentally compared ReMUS with the state-of-the-art MUS enumeration algorithm MARCO in the SAT and SMT domains. The results show that the tendency to minimise the number of performed satisfiability checks leads to a~significant improvement over the state-of-the-art.

\bibliography{refs_concise}

\end{document}